\documentclass[acmsmall]{acmart}
\AtBeginDocument{%
  }

\setcopyright{acmlicensed}
\copyrightyear{2024}
\acmYear{2024}
\acmDOI{XXXXXXX.XXXXXXX}

\acmConference[Conference acronym 'XX]{Make sure to enter the correct
  conference title from your rights confirmation emai}{June 03--05,
  2024}{Woodstock, NY}
\acmISBN{978-1-4503-XXXX-X/18/06}

\begin{document}

\title{Data Collection and Labeling Techniques for Machine Learning}

\date{\today}

\author{Qianyu Huang}
\email{qianyuhuang19@gmail.com
}
\affiliation{%
  \institution{National Taiwan University of Science and Technology}
  \city{Taipei}
  \country{Taiwan}
}

\author{Tongfang Zhao}
\affiliation{%
  \institution{National Taiwan University of Science and Technology}
  \city{Taipei}
  \country{Taiwan}}

\begin{abstract}
Data collection and labeling are critical bottlenecks in the deployment of machine learning applications. With the increasing complexity and diversity of applications, the need for efficient and scalable data collection and labeling techniques has become paramount. This paper provides a review of the state-of-the-art methods in data collection, data labeling, and the improvement of existing data and models. By integrating perspectives from both the machine learning and data management communities, we aim to provide a holistic view of the current landscape and identify future research directions.
\end{abstract}

\maketitle

\section{Introduction}
The realm of machine learning (ML) has undergone a period of unprecedented growth in recent years. This remarkable advancement can be attributed to two key factors: the exponential rise in computational power and the ever-increasing availability of vast datasets~\cite{lecun2015deep, chen2014big, chen2019deep}. However, the very foundation upon which this progress rests – data collection and labeling – presents significant challenges that can hinder the efficacy and ethical implementation of ML models~\cite{roh2019survey, jo2020lessons, brodley2012challenges, wu2023ground, tizghadam2019machine}. This review paper delves into the intricate world of data collection and labeling for machine learning, drawing upon insights from both the data management and machine learning communities.

The transformative potential of machine learning is evident across a multitude of domains. From revolutionizing healthcare with disease diagnosis and personalized medicine~\cite{raparthi2020deep} to powering self-driving cars~\cite{badue2021self} and optimizing logistics in supply chains~\cite{odimarha2024machine}, ML algorithms are rapidly reshaping our world. At the heart of these advancements lies the ability of ML models to learn from data, identify patterns, and make predictions based on the information they have been exposed to. The quality and quantity of data used to train these models are paramount to their success. High-quality, diverse, and well-labeled data are essential for building robust and generalizable ML models that can perform effectively in real-world scenarios~\cite{han2020survey, 10.1145/3491232}.

However, the process of collecting and labeling data for machine learning is far from straightforward. The sheer volume of data required to train complex models can be daunting, and the task of meticulously labeling each data point can be incredibly time-consuming and expensive.  Furthermore, ethical considerations regarding data privacy and potential biases within datasets pose significant challenges that must be addressed to ensure responsible and trustworthy development of ML applications.

This review paper aims to provide a comprehensive analysis of the various techniques and methodologies employed in data collection and labeling for machine learning. By integrating insights from both data management and ML perspectives, we will explore the intricacies of this critical process. We will delve into the diverse techniques used to gather data across different data types. We will then analyze the various approaches employed to label this raw data, including manual labeling, automated labeling techniques, and the potential of crowdsourcing platforms.

Furthermore, we will explore how data management practices like data cleaning, standardization, and versioning can be seamlessly integrated into data collection and labeling workflows to enhance efficiency and ensure data quality.
Finally, this review paper will explore future directions for research and development in data collection and labeling for machine learning. With the field constantly evolving, the need for innovative strategies to gather and label data will continue to grow.  We will examine the potential of semi-supervised and unsupervised learning techniques to reduce reliance on labeled data, as well as delve into the development of robust methods for bias detection and mitigation in datasets.

By providing a detailed analysis of the current state of data collection and labeling for machine learning, drawing upon insights from both data management and ML perspectives, and outlining future research directions, this review paper aims to serve as a valuable resource for researchers, practitioners, and anyone interested in understanding the foundations upon which the transformative power of machine learning is built.

The structure of this review is as follows: Section 2 discusses data acquisition techniques, including data discovery, augmentation, and generation. Section 3 delves into data labeling methods, covering existing labels, crowd-based techniques, and weak supervision. Finally, Section 4 discusses future research directions and concludes the review.

\section{Data Collection Techniques}
Data collection can be broadly categorized into three approaches: data acquisition, data labeling, and the improvement of existing data or models.

\subsection{Data Acquisition}
Data acquisition involves discovering, augmenting, or generating datasets.

\subsubsection{Data Discovery}
Data discovery serves as the cornerstone of successful machine learning endeavors. It entails the systematic process of identifying, locating, and accessing datasets relevant to a specific machine learning task. This retrieved data acts as the raw material that fuels the learning process of a machine learning algorithm, ultimately influencing its effectiveness in achieving the desired outcome.

\noindent\textbf{Challenges and Traditional Approaches}: The data landscape can be vast and heterogeneous, encompassing structured datasets stored in relational databases to unstructured data residing in social media platforms or sensor networks. Traditionally, data discovery has been a cumbersome and time-consuming process, often involving:
\begin{itemize}
    \item Manual exploration of data repositories.
    \item Reliance on domain expertise to locate relevant datasets.
    \item Limited interoperability between data sources.
\end{itemize}

\paragraph{Collaborative Analysis}
Collaborative data analysis platforms like DataHub~\cite{bhardwaj2014datahub} address the growing need for efficient data management and analysis in multi-user environments.  DataHub offers a centralized repository for data hosting, sharing, and manipulation.  Inspired by the familiar Git version control system, DataHub facilitates dataset versioning, allowing users to track changes, revert to previous iterations, and seamlessly merge work from different teams on the same data. This eliminates data silos and version conflicts, fostering a collaborative workspace.  Furthermore, DataHub provides a hosted platform with integrated functionalities for data discovery (search), quality improvement (cleaning), and integration from diverse sources. Additionally, built-in visualization tools enable the transformation of data insights into clear and actionable visuals.  By centralizing data management, version control, and offering a suite of data manipulation tools, DataHub empowers researchers and analysts to work efficiently, ensuring access to high-quality, trusted data for robust and collaborative research endeavors.

\paragraph{Web-based Systems}
Web-based systems have democratized data exploration and sharing. Platforms like Google Fusion Tables~\cite{gonzalez2010google} empower users to upload structured data and leverage a suite of visual analysis tools for filtering, aggregation, and exploration. Fusion Tables further enhance discoverability by facilitating dataset publication on the web, making them searchable through search engines. Additionally, data marketplaces like CKAN~\cite{winn2013open}, Quandl~\cite{QuandlAl73:online}, and DataMarket~\cite{DataMark55:online} provide a dedicated infrastructure for dataset sharing and discovery. These platforms connect data providers with potential users, fostering collaboration and accelerating research progress. By simplifying data access and visualization, web-based systems play a crucial role in promoting open data practices and driving data-driven innovation across various disciplines.

\paragraph{Integrated Platforms}
Integrated platforms like Kaggle~\cite{KaggleYo44:online} have emerged as a cornerstone for fostering collaboration and innovation within the data science community. These platforms combine the power of web-based data sharing with collaborative analysis tools, creating a dynamic environment for practitioners of all skill levels.

One of Kaggle's core functionalities lies in hosting data science competitions. These competitions provide participants with access to shared datasets, allowing them to experiment with their own techniques and approaches to solving real-world problems.  The competitive nature of these events incentivizes participants to push the boundaries of their knowledge and develop novel methodologies. Additionally, Kaggle fosters knowledge exchange through leaderboards and discussion forums, enabling participants to learn from each other's successes and failures.

Beyond competitions, Kaggle offers a vast repository of publicly available datasets, encompassing a diverse range of domains. This eliminates the initial hurdle of data acquisition for many machine learning projects, allowing practitioners to focus on analysis and model development.  Furthermore, Kaggle facilitates the sharing of code snippets and notebooks, enabling users to not only access data but also learn from the approaches employed by others. This fosters a culture of open-source development and accelerates progress within the field.

\subsubsection{Data Searching in Data Lakes and on the Web}
Data searching systems are designed to help users find relevant datasets within large repositories, such as corporate data lakes or the web.

\paragraph{Data Lakes}
Data lakes have emerged as a powerful paradigm for storing and managing massive volumes of data in its native format. Unlike traditional data warehouses, which require pre-defined schema, data lakes offer a flexible and scalable solution  by accommodating structured, semi-structured, and unstructured data~\cite{john2017data, vcuvs2022data, gupta2018practical}. This versatility is exemplified by prominent offerings like IBM's data lake~\cite{panwar2022blockchain} and Google's GOODS~\cite{halevy2016goods}, which address the challenge of efficiently indexing and searching across vast datasets.

The core functionality of a data lake lies in its ability to catalog metadata from diverse storage systems~\cite{sawadogo2021data}. This metadata acts as a comprehensive index, enabling users to conduct keyword searches and explore detailed profiles of datasets within the data lake. This empowers researchers and data analysts to discover relevant datasets with greater ease, fostering a data-driven approach to problem-solving. Furthermore, data lakes often support expressive queries, allowing users to search based on specific criteria beyond basic keywords~\cite{mami2019uniform}. This advanced search capability facilitates more granular exploration and retrieval of relevant data subsets.

In essence, data lakes act as a central hub for data discovery, streamlining the process of finding and utilizing valuable datasets. This enhanced accessibility fosters collaboration and innovation within research communities, as researchers can leverage the collective data resources stored within the lake. By democratizing access to data, data lakes play a crucial role in accelerating scientific progress and uncovering new knowledge.

\paragraph{Web-based Data Extraction}
Web-based data extraction systems  play a crucial role in harvesting structured information from the vast amount of unstructured data residing on the web. One such system, WebTables~\cite{cafarella2018ten}, automates the process of extracting valuable data from HTML tables. This technology goes beyond simply copying and pasting table content. WebTables delves deeper, identifying relational tables within web pages and extracting both the schema (column headers defining data types) and the tuples (individual rows containing data points). This capability unlocks the potential to transform raw HTML tables into a format readily usable for further analysis and integration with other datasets.

The success of WebTables has spurred advancements in related data extraction techniques. Researchers have extended the core functionalities to encompass the extraction of vertical tables, which present data in a column-oriented format, often used for product listings or specifications~\cite{stonebraker2018c}. Additionally, the ability to extract information from list structures and knowledge bases has been incorporated, further enriching the range of web content that can be processed~\cite{ritze2016profiling}. These advancements demonstrate the adaptability and potential of web-based data extraction systems to evolve alongside the ever-changing structure of web content.

\subsubsection{Data Augmentation}
Data augmentation techniques enhance existing datasets by adding more external data or integrating multiple sources.

\paragraph{Deriving Latent Semantics}
Latent semantic analysis (LSA) can be significantly enhanced by leveraging word embedding techniques such as Word2Vec~\cite{church2017word2vec}, GloVe~\cite{pennington2014glove}, and Doc2Vec~\cite{trucscua2019efficiency}. These techniques go beyond simple word co-occurrence statistics traditionally used in LSA, instead capturing the nuanced semantic relationships between words, entities, and even documents within a corpus. This is achieved by training the embedding models on large text datasets, where words appearing in similar contexts are gradually positioned closer together in a high-dimensional vector space. As a result, these embeddings effectively encode the linguistic context of words, allowing them to represent not just the literal meaning, but also the underlying concepts and relationships.

This newfound richness in feature representation translates to significant advantages for machine learning tasks that rely on LSA. In traditional LSA, documents and terms are represented as sparse, high-dimensional vectors where most elements hold a value of zero~\cite{wang2017machine}. Embedding-based LSA, on the other hand, utilizes the dense, low-dimensional vectors learned from embedding models. These denser representations capture the semantic relationships more effectively, leading to improved performance in tasks like document retrieval, topic modeling, and text classification. For instance, a document discussing "cellular phones" can be semantically linked to documents about "mobile devices" or "smartphones" through the shared concept of mobile communication, even though the literal terms themselves might not be present.

By incorporating word embeddings, LSA gains the ability to handle synonyms, polysemy (words with multiple meanings), and even out-of-vocabulary terms that were not present in the training data. This robustness allows LSA to perform more accurate semantic analysis and opens doors for various applications in natural language processing.

\paragraph{Entity Augmentation}
Entity augmentation techniques address the challenge of incomplete datasets by enriching them with additional information extracted from external sources~\cite{cao2020open}. This approach is particularly valuable when dealing with datasets containing missing values or attributes. Two prominent examples of such techniques are Octopus~\cite{sun2019gathering} and InfoGather~\cite{yakout2012infogather}. These methods leverage the vast amount of information available in web tables to fill these gaps.

The core concept behind entity augmentation lies in the ability to match and integrate relevant data points from multiple web tables. This process typically involves two key steps: schema matching and entity resolution. Schema matching aims to identify semantically equivalent attributes across different web tables. For instance, the attribute "City" in one table might correspond to the attribute "Location" in another. Entity resolution, on the other hand, focuses on linking mentions of the same entity across various tables. This ensures that the information extracted refers to the same real-world entity being represented within the dataset.

By effectively employing schema matching and entity resolution, entity augmentation techniques can significantly enhance the completeness and accuracy of datasets. This allows for more comprehensive analysis and improved performance in tasks that rely on rich and informative data. However, it is crucial to acknowledge potential limitations. The quality and relevance of the extracted information heavily depend on the accuracy of the underlying web tables themselves. Additionally, ensuring the coherence and consistency of the augmented data necessitates careful consideration during the integration process.

\paragraph{Data Integration}
Data integration is a crucial process that involves the combination of datasets originating from various sources to create a single, unified dataset~\cite{hamid2009data, li2021deep, 10.1145/3431816, 10.1145/3318464.3389743}. This unified dataset is essential for a comprehensive analysis, as it consolidates information from multiple origins to provide a more complete view of the data landscape. The process of data integration can be complex, requiring sophisticated techniques to ensure that the resulting dataset is both accurate and efficient to use.

The importance of data integration lies in its ability to merge disparate data sources, which may include databases, data warehouses, and external data feeds, into a coherent whole. This is particularly critical in fields such as healthcare~\cite{JAYARATNE2019996}, finance~\cite{10.1007/978-3-319-07443-6_40}, and scientific research~\cite{10.14778/3229863.3240491}, where data is often scattered across various platforms and formats. By integrating these diverse data sources, organizations can achieve a holistic view that supports better decision-making and insights.

One of the primary challenges in data integration is dealing with the heterogeneity of data sources. Data from different sources may vary in terms of format, structure, and semantics. For instance, one database might store dates in the format "YYYY-MM-DD," while another uses "DD/MM/YYYY." Similarly, the same entity might be represented differently across datasets, such as "John Smith" in one source and "J. Smith" in another. To address these issues, data integration processes often involve data cleaning and transformation steps to standardize the data.

Data cleaning involves identifying and correcting errors or inconsistencies in the data~\cite{10.1145/2882903.2912574}. This can include tasks such as removing duplicate records, filling in missing values, and correcting typos~\cite{karlaš2020nearest, 7474370, 6816736, 9835412}. Data transformation, on the other hand, involves converting data from one format or structure to another to ensure compatibility between datasets. This might include normalizing data to a common scale, converting data types, or restructuring data to match a desired schema.

Schema matching is another critical aspect of data integration~\cite{1410106}. This involves aligning the schemas (i.e., the structure and organization) of different datasets to ensure that corresponding data elements are correctly mapped to each other. Schema matching can be particularly challenging when dealing with complex or poorly documented datasets. Automated schema matching techniques, such as those based on machine learning or heuristic algorithms, can help to streamline this process, but manual intervention is often required to resolve ambiguities and ensure accuracy.

Once data has been cleaned, transformed, and schema-matched, it must be merged into a single dataset. This merging process can involve various strategies, such as union, join, or aggregation operations, depending on the nature of the data and the desired outcome. For example, a union operation might be used to combine rows from two datasets with the same schema, while a join operation might be used to combine data based on a common key, such as a customer ID.

Data integration also involves addressing issues related to data quality and consistency. Ensuring that the integrated dataset is accurate and reliable is paramount, as errors or inconsistencies can lead to incorrect conclusions and decisions. Techniques such as data validation, which checks for correctness and completeness, and data reconciliation, which resolves discrepancies between datasets, are essential components of the data integration process.

Scalability is another important consideration in data integration. As the volume of data continues to grow, integration processes must be able to handle large-scale datasets efficiently. This requires the use of scalable algorithms and technologies, such as distributed computing frameworks and parallel processing techniques, to ensure that data integration tasks can be performed in a timely manner.

In recent years, advances in artificial intelligence and machine learning have significantly enhanced data integration techniques. Machine learning algorithms can be used to automate various aspects of the integration process, such as schema matching, data cleaning, and entity resolution. These algorithms can learn from examples and improve over time, making them particularly effective for dealing with complex and heterogeneous data sources.

Deep learning, a subset of machine learning, has also shown promise in the field of data integration. Deep learning models, such as neural networks, can be trained to recognize patterns and relationships in data, enabling them to perform tasks such as feature extraction and data transformation with high accuracy. These models can be particularly useful for integrating unstructured data, such as text or images, with structured data.

Despite these advancements, data integration remains a challenging and resource-intensive task. It requires careful planning and execution to ensure that the resulting dataset is accurate, consistent, and useful for analysis. Moreover, the integration process must be continuously monitored and maintained, as data sources and requirements can change over time.

\subsubsection{Data Generation}
In many research scenarios, the available datasets may not meet the specific needs of a study, either due to limitations in size, diversity, or specificity. To address these gaps, researchers often turn to alternative methods for data generation. Two prominent approaches in this context are crowdsourcing and synthetic data generation, each offering unique advantages and methodologies for creating new datasets that can significantly enhance the scope and depth of research.

\paragraph{Crowdsourcing}
Crowdsourcing has emerged as a powerful tool for data generation, leveraging the collective efforts of a large number of individuals. Platforms such as Amazon Mechanical Turk~\cite{AmazonMe93:online} facilitate this process by providing a marketplace where researchers can post tasks that are then completed by human workers. These tasks can range from simple data entry to more complex activities such as image annotation, sentiment analysis, and even creative tasks like writing or drawing.

The use of crowdsourcing allows for the rapid collection of large volumes of data, which can be particularly useful in fields where human judgment and perception are critical. For instance, in natural language processing, crowdsourced workers can be employed to label datasets for training machine learning models. Similarly, in computer vision, workers can annotate images to help improve object detection algorithms.

To ensure the accuracy and reliability of the collected data, various quality control measures are implemented. These may include redundancy, where multiple workers complete the same task to cross-verify results, and the use of gold standard questions, where known answers are interspersed with the tasks to monitor workerperformance. Advanced quality control frameworks, such as TurKit~\cite{10.1145/1866029.1866040}, AutoMan~\cite{10.1145/2398857.2384663}, and Deco~\cite{10.1145/2396761.2398421}, have been developed to streamline these processes. TurKit, for example, allows for iterative refinement of tasks, enabling researchers to dynamically adjust task parameters based on preliminary results. AutoMan integrates human computation seamlessly with traditional programming, automatically managing task assignments and quality checks. Deco provides a declarative approach to crowdsourcing, allowing researchers to specify what data they need and letting the system handle the logistics of data collection and validation.

By harnessing the power of crowdsourcing, researchers can generate high-quality datasets tailored to their specific needs, often at a fraction of the cost and time required for traditional data collection methods. This approach is particularly valuable in rapidly evolving fields where up-to-date and context-specific data is crucial.

\paragraph{Synthetic Data Generation}
In contrast to crowdsourcing, synthetic data generation leverages computational techniques to create artificial datasets that mimic the properties of real-world data. One of the most advanced and widely used methods in this domain is the use of Generative Adversarial Networks (GANs) \cite{10.1145/3422622, ARMANIOUS2020101684, pmlr-v157-zhao21a}. GANs consist of two neural networks—a generative network and a discriminative network—that are trained simultaneously through a process of adversarial learning. The generative network attempts to produce realistic data samples, while the discriminative network evaluates these samples against real data, providing feedback that helps the generative network improve.

The application of GANs has revolutionized synthetic data generation, enabling the creation of large-scale datasets with remarkable fidelity. For instance, GANs have been used to generate synthetic images that are nearly indistinguishable from real photographs, which is particularly useful in fields like computer vision and graphics. MedGAN \cite{ARMANIOUS2020101684} and TableGAN \cite{pmlr-v157-zhao21a} extend these capabilities to medical and tabular data, respectively, allowing for the creation of synthetic health records and structured data that can be used for research without compromising patient privacy or data security.

Synthetic data generation offers several key advantages. It allows researchers to create datasets that are perfectly balanced and free from the biases that often plague real-world data. It also enables the generation of rare or edge-case scenarios that might be underrepresented in existing datasets, thereby improving the robustness and generalizability of machine learning models. Additionally, synthetic data can be used to augment real datasets, providing additional training examples that enhance model performance.

However, the use of synthetic data also presents challenges. Ensuring that synthetic data accurately reflects the complexities and nuances of real-world data is a non-trivial task. Overfitting to the synthetic data or failing to capture critical variations can lead to models that perform poorly in real-world applications. Therefore, rigorous validation and testing against real data are essential to ensure the utility and reliability of synthetic datasets.

In summary, both crowdsourcing and synthetic data generation offer powerful tools for creating new datasets when existing ones are insufficient. Crowdsourcing leverages human intelligence to gather and annotate data, while synthetic data generation uses advanced computational techniques to produce artificial data that mimics real-world properties. By employing these methods, researchers can overcome data limitations, enhance the quality and diversity of their datasets, and ultimately drive forward the frontiers of knowledge in their respective fields.

\section{Data Labeling Techniques}
Data labeling is crucial for supervised machine learning. It can be achieved through crowd-based methods and and weak supervision.

\subsection{Crowd-based Techniques}
Crowd-based techniques are a set of methodologies that leverage the collective intelligence and efforts of human workers to label data. These techniques can be broadly categorized into two main types: active learning and general crowdsourcing methods. Both approaches aim to enhance the quality and efficiency of data labeling processes, which are crucial for training machine learning models.

\subsubsection{Active Learning}
Active learning is a subset of machine learning in which the algorithm selectively chooses the data from which it learns. The primary objective of active learning is to identify the most informative examples for labeling, thereby reducing the amount of labeled data required while maintaining or improving model performance. Several strategies are employed within active learning to achieve this goal:

Uncertainty Sampling: This technique involves selecting data points for which the model is least certain about the correct label. By focusing on these uncertain instances, the model can quickly improve its understanding of the data distribution \cite{JANSSEN2013123}.

Query-by-Committee: In this approach, multiple models (the committee) are trained on the same dataset, and the instances on which these models disagree the most are selected for labeling. This method leverages the diversity of opinions among the models to identify the most informative examples \cite{10.1145/130385.130417}.

Density Weighting: This strategy selects examples not only based on uncertainty but also on the density of the data points in the feature space. By choosing examples from dense regions, the model can learn more about the underlying structure of the data, which can be particularly useful in cases where the data distribution is complex \cite{10.1145/3534678.3539476}.

Moreover, decision-theoretic approaches in active learning focus on optimizing the selection of examples based on specific performance metrics of the model. These approaches consider the expected improvement in model performance that labeling a particular example would bring, thereby making the labeling process more efficient and targeted \cite{kapoor2007selective}.

\subsubsection{Crowdsourcing}
Crowdsourcing is a technique that involves distributing labeling tasks to a large pool of human workers, typically through online platforms. This method harnesses the power of the crowd to perform tasks that are difficult or infeasible for a single individual or a small group to handle. Several key techniques and platforms have been developed to enhance the effectiveness and reliability of crowdsourcing for data labeling~\cite{10.1145/3025453.3026044}:

Revolt: This platform provides a framework for repeated labeling and aggregation, ensuring that the final labels are of high quality. Revolt incorporates mechanisms to detect and mitigate the effects of low-quality work by individual workers, thereby improving the overall accuracy of the labeled data \cite{10.1145/3025453.3026044}.

CrowdER: CrowdER is designed to handle entity resolution tasks, which involve identifying and merging records that refer to the same entity. This platform uses crowdsourcing to resolve ambiguities and discrepancies in data, leveraging human judgment to achieve high accuracy in entity matching \cite{wang2012crowder}.

Qurk: Qurk is a query processing system that integrates crowdsourcing into database operations. It provides interfaces and mechanisms to streamline the process of assigning labeling tasks to human workers, ensuring that the tasks are completed efficiently and accurately. Qurk also includes features to manage the quality of the work performed by the crowd, such as redundancy and aggregation techniques to verify the correctness of the labels \cite{10.1145/1989323.1989486}.

In summary, crowd-based techniques, encompassing both active learning and crowdsourcing, play a vital role in the data labeling process. Active learning focuses on selecting the most informative examples for labeling, thereby reducing the amount of labeled data needed. On the other hand, crowdsourcing leverages the collective efforts of human workers to label large datasets accurately and efficiently. Both approaches contribute significantly to the development of robust and reliable machine learning models by ensuring that the training data is of high quality.

\subsection{Weak Supervision}
Weak supervision is an innovative approach in machine learning that focuses on generating large quantities of labeled data using less precise, but more scalable, labeling methods. This approach is particularly valuable when obtaining a large volume of high-quality labeled data is impractical due to time, cost, or resource constraints. Weak supervision methods typically involve the use of labeling functions, heuristics, and other automated techniques to produce weak labels, which can then be refined and used to train machine learning models. The core idea is to leverage various sources of weak supervision to create a sufficiently large dataset that can approximate the quality of a fully supervised dataset.

\subsubsection{Data Programming}
Data programming is one of the foundational techniques in weak supervision. It is implemented in systems such as Snorkel \cite{10.1145/3035918.3056442, Wu_2021}, which has gained significant attention for its ability to generate weak labels efficiently. The process involves the following steps~\cite{zhang2022survey}:

Labeling Functions: At the heart of data programming are labeling functions, which are simple, user-defined rules or heuristics that assign labels to data points. These functions can be based on domain knowledge, pattern matching, or other automated methods. Each labeling function may be noisy and imperfect, but collectively, they can provide valuable signals about the data.

Generative Model: The weak labels generated by the labeling functions are combined into a generative model. This model estimates the accuracy and correlations of the labeling functions, allowing it to better understand the noise and biases inherent in the weak labels. By modeling these aspects, the generative model can produce a set of probabilistic labels that are more accurate than the individual outputs of the labeling functions.

Discriminative Model: The probabilistic labels generated by the generative model are then used to train a discriminative model. This model, typically a machine learning classifier, learns to predict the target labels from the input data. The discriminative model benefits from the large volume of weakly labeled data, enabling it to generalize well to new, unseen data.

Data programming thus leverages the strengths of both automated and manual labeling techniques, providing a scalable solution for generating labeled datasets. By combining multiple weak signals, it can produce high-quality labels that are suitable for training robust machine learning models.

\subsubsection{Fact Extraction}
Fact extraction is another critical technique in weak supervision, particularly useful for constructing knowledge bases and generating seed labels for distant supervision. Systems like NELL (Never-Ending Language Learning) \cite{10.1145/3191513} and Knowledge Vault \cite{10.1145/2623330.2623623} exemplify the use of fact extraction in weak supervision.

Structured Information Extraction: Fact extraction involves extracting structured information from unstructured web data, such as text documents, web pages, and other sources. Techniques used for fact extraction include natural language processing (NLP), pattern matching, and machine learning. The goal is to identify and extract entities, relationships, and attributes that can be used to construct a knowledge base.

Knowledge Bases: The extracted facts are organized into knowledge bases, which are structured repositories of information. These knowledge bases can be used for various applications, such as question answering, information retrieval, and semantic search. In the context of weak supervision, the facts in the knowledge base serve as seed labels for distant supervision.

Distant Supervision: Distant supervision is a technique where the seed labels generated from the knowledge base are used to label large amounts of data automatically. For example, if a knowledge base contains information about the relationships between entities, these relationships can be used to label sentences in a text corpus that mention the same entities. Although the labels generated through distant supervision may be noisy, they provide a valuable starting point for training machine learning models.

Systems like NELL and Knowledge Vault continuously extract and update facts from the web, enabling them to generate an ever-growing set of weak labels. These labels can be used to train models for various tasks, such as entity recognition, relation extraction, and more. By leveraging the vast amount of information available on the web, fact extraction techniques provide a scalable and efficient way to generate weak labels for machine learning.

In summary, weak supervision methods, including data programming and fact extraction, offer powerful solutions for generating large quantities of labeled data. Data programming combines multiple labeling functions into a generative model, which is then used to train a discriminative model, leveraging both automated and manual labeling techniques. Fact extraction techniques, on the other hand, extract structured information from unstructured data to construct knowledge bases, which can be used as seed labels for distant supervision. Both approaches enable the creation of high-quality labeled datasets, facilitating the training of robust machine learning models even in scenarios where obtaining fully supervised data is challenging.

\section{Discussion and Future Directions}
Despite the significant advancements in data collection and labeling techniques, several challenges persist that need to be addressed to further enhance the efficiency and effectiveness of machine learning workflows. These challenges include evaluating the sufficiency and quality of collected data, integrating various techniques into a cohesive system, addressing performance trade-offs, and developing comprehensive end-to-end solutions. Addressing these areas will be critical for advancing the field and enabling the development of more robust and accurate machine learning models.

\subsection{Evaluating Data Quality}
One of the most pressing challenges in the field of machine learning is determining whether the data collected is of high quality and sufficient for training accurate models~\cite{9417095}. The quality of the data directly impacts the performance of the machine learning models, making it crucial to develop robust techniques for evaluating data quality. Current methods for data evaluation often rely on manual inspection or simple statistical measures, which can be time-consuming and may not always provide a complete picture of the data's quality.

Future research should focus on developing more sophisticated techniques for evaluating data quality. These techniques could include automated methods for detecting and correcting errors, identifying biases, and assessing the representativeness of the data. Additionally, methods for selecting the most relevant datasets for a given task need to be refined. This could involve developing algorithms that can automatically identify and prioritize the most informative data points, similar to the principles used in active learning.

Moreover, the development of standardized metrics and benchmarks for data quality evaluation would be beneficial. Such metrics would enable practitioners to objectively assess the quality of their datasets and make more informed decisions about data collection and labeling strategies. Ultimately, improving data quality evaluation techniques will lead to the creation of more reliable and accurate machine learning models.

\subsection{Integrating Techniques}
Integrating various data collection and labeling techniques into a cohesive workflow is essential for maximizing efficiency and effectiveness in the data preparation process. Currently, many machine learning workflows involve a fragmented approach, where different techniques are applied in isolation. This can lead to inefficiencies and inconsistencies in the data preparation process, ultimately affecting the performance of the final model.

Developing frameworks that combine data acquisition, labeling, and improvement methods into a unified system will help streamline the data preparation process. Such frameworks should be designed to facilitate seamless integration of different techniques, allowing practitioners to leverage the strengths of each method. For example, a unified framework could combine active learning with crowdsourcing, using active learning to identify the most informative examples and crowdsourcing to obtain high-quality labels for those examples.

Additionally, these frameworks should be flexible and adaptable, allowing practitioners to customize their workflows based on the specific requirements of their tasks. This could involve developing modular systems that enable users to easily add or remove components, as well as providing tools for monitoring and optimizing the data preparation process. By integrating various techniques into a cohesive workflow, practitioners can improve the efficiency and effectiveness of their data preparation efforts, leading to better model performance.

\subsection{Performance Trade-offs}
Balancing the trade-offs between accuracy and scalability is a key consideration in data collection and labeling. Different techniques often come with their own set of trade-offs, and understanding the impact of these trade-offs on model performance and computational overhead is crucial for making informed decisions.

For example, active learning techniques can significantly reduce the amount of labeled data needed to train a model, but they may also require more computational resources to identify the most informative examples~\cite{NEURIPS2019_95323660}. Similarly, crowdsourcing can provide high-quality labels, but it may be costly and time-consuming to manage a large pool of human workers~\cite{10.1145/3289600.3291035}. On the other hand, weak supervision methods can generate large quantities of labeled data quickly and cheaply, but the quality of the labels may be lower compared to fully supervised methods.

Future research should focus on developing methods for quantifying and balancing these trade-offs. This could involve creating models that can predict the impact of different data collection and labeling techniques on model performance and computational overhead. Additionally, developing optimization algorithms that can automatically balance these trade-offs based on the specific constraints and requirements of a given task would be beneficial.

Understanding the performance trade-offs associated with different techniques will enable practitioners to make more informed decisions about their data collection and labeling strategies. This, in turn, will lead to the development of more efficient and effective machine learning workflows.

\subsection{End-to-end Solutions}
Developing end-to-end solutions that encompass data acquisition, labeling, and improvement techniques will be crucial for advancing the field of machine learning. These solutions should be designed to be flexible, scalable, and easy to use, enabling practitioners to efficiently prepare data for machine learning tasks.

End-to-end solutions should integrate various data collection and labeling techniques into a single, cohesive system. This could involve developing platforms that provide tools for data acquisition, labeling, and quality evaluation, as well as methods for improving the quality of the labeled data. Such platforms should be designed to facilitate seamless integration of different techniques, allowing practitioners to leverage the strengths of each method.

Additionally, end-to-end solutions should be scalable, enabling practitioners to handle large volumes of data efficiently. This could involve developing distributed systems that can process and label data in parallel, as well as tools for managing and monitoring the data preparation process. Scalability is particularly important in the context of big data, where the volume of data can be overwhelming.

Flexibility is another key consideration for end-to-end solutions. These solutions should be adaptable to different types of data and tasks, allowing practitioners to customize their workflows based on their specific requirements. This could involve providing modular components that can be easily added or removed, as well as tools for configuring and optimizing the data preparation process.

Ease of use is also critical for the adoption of end-to-end solutions. These solutions should provide user-friendly interfaces and tools that enable practitioners to easily manage and monitor their data preparation efforts. This could involve developing visual interfaces for data labeling and quality evaluation, as well as providing tools for automating repetitive tasks.

In summary, despite the advancements in data collection and labeling techniques, several challenges remain that need to be addressed to further enhance the efficiency and effectiveness of machine learning workflows. Evaluating the sufficiency and quality of collected data, integrating various techniques into a cohesive system, addressing performance trade-offs, and developing comprehensive end-to-end solutions are critical areas for future research. Addressing these challenges will be essential for advancing the field and enabling the development of more robust and accurate machine learning models.

\section{Conclusion}
This review provides a comprehensive overview of data collection and labeling techniques for machine learning, integrating insights from both the machine learning and data management communities. By addressing the challenges and identifying future research directions, we hope to guide researchers and practitioners in developing more efficient and scalable solutions for data-driven machine learning applications.

\bibliographystyle{IEEEtran}
\bibliography{references}

\begin{thebibliography}{10}
\providecommand{\url}[1]{#1}
\csname url@samestyle\endcsname
\providecommand{\newblock}{\relax}
\providecommand{\bibinfo}[2]{#2}
\providecommand{\BIBentrySTDinterwordspacing}{\spaceskip=0pt\relax}
\providecommand{\BIBentryALTinterwordstretchfactor}{4}
\providecommand{\BIBentryALTinterwordspacing}{\spaceskip=\fontdimen2\font plus
\BIBentryALTinterwordstretchfactor\fontdimen3\font minus \fontdimen4\font\relax}
\providecommand{\BIBforeignlanguage}[2]{{%
\expandafter\ifx\csname l@#1\endcsname\relax
\typeout{** WARNING: IEEEtran.bst: No hyphenation pattern has been}%
\typeout{** loaded for the language `#1'. Using the pattern for}%
\typeout{** the default language instead.}%
\else
\language=\csname l@#1\endcsname
\fi
#2}}
\providecommand{\BIBdecl}{\relax}
\BIBdecl

\bibitem{lecun2015deep}
Y.~LeCun, Y.~Bengio, and G.~Hinton, ``Deep learning,'' \emph{nature}, vol. 521, no. 7553, pp. 436--444, 2015.

\bibitem{chen2014big}
X.-W. Chen and X.~Lin, ``Big data deep learning: challenges and perspectives,'' \emph{IEEE access}, vol.~2, pp. 514--525, 2014.

\bibitem{chen2019deep}
J.~Chen and X.~Ran, ``Deep learning with edge computing: A review,'' \emph{Proceedings of the IEEE}, vol. 107, no.~8, pp. 1655--1674, 2019.

\bibitem{roh2019survey}
Y.~Roh, G.~Heo, and S.~E. Whang, ``A survey on data collection for machine learning: a big data-ai integration perspective,'' \emph{IEEE Transactions on Knowledge and Data Engineering}, vol.~33, no.~4, pp. 1328--1347, 2019.

\bibitem{jo2020lessons}
E.~S. Jo and T.~Gebru, ``Lessons from archives: Strategies for collecting sociocultural data in machine learning,'' in \emph{Proceedings of the 2020 conference on fairness, accountability, and transparency}, 2020, pp. 306--316.

\bibitem{brodley2012challenges}
C.~E. Brodley, U.~Rebbapragada, K.~Small, and B.~Wallace, ``Challenges and opportunities in applied machine learning,'' \emph{Ai Magazine}, vol.~33, no.~1, pp. 11--24, 2012.

\bibitem{wu2023ground}
R.~Wu, A.~Bendeck, X.~Chu, and Y.~He, ``Ground truth inference for weakly supervised entity matching,'' \emph{Proceedings of the ACM on Management of Data}, vol.~1, no.~1, pp. 1--28, 2023.

\bibitem{tizghadam2019machine}
A.~Tizghadam, H.~Khazaei, M.~H. Moghaddam, and Y.~Hassan, ``Machine learning in transportation.'' \emph{Journal of Advanced Transportation}, 2019.

\bibitem{raparthi2020deep}
M.~Raparthi, ``Deep learning for personalized medicine-enhancing precision health with ai,'' \emph{Journal of Science \& Technology}, vol.~1, no.~1, pp. 82--90, 2020.

\bibitem{badue2021self}
C.~Badue, R.~Guidolini, R.~V. Carneiro, P.~Azevedo, V.~B. Cardoso, A.~Forechi, L.~Jesus, R.~Berriel, T.~M. Paixao, F.~Mutz \emph{et~al.}, ``Self-driving cars: A survey,'' \emph{Expert systems with applications}, vol. 165, p. 113816, 2021.

\bibitem{odimarha2024machine}
A.~C. Odimarha, S.~A. Ayodeji, and E.~A. Abaku, ``Machine learning's influence on supply chain and logistics optimization in the oil and gas sector: a comprehensive analysis,'' \emph{Computer Science \& IT Research Journal}, vol.~5, no.~3, pp. 725--740, 2024.

\bibitem{han2020survey}
B.~Han, Q.~Yao, T.~Liu, G.~Niu, I.~W. Tsang, J.~T. Kwok, and M.~Sugiyama, ``A survey of label-noise representation learning: Past, present and future,'' \emph{arXiv preprint arXiv:2011.04406}, 2020.

\bibitem{10.1145/3491232}
\BIBentryALTinterwordspacing
R.~Wu, N.~Das, S.~Chaba, S.~Gandhi, D.~H. Chau, and X.~Chu, ``A cluster-then-label approach for few-shot learning with application to automatic image data labeling,'' \emph{J. Data and Information Quality}, vol.~14, no.~3, may 2022. [Online]. Available: \url{https://doi.org/10.1145/3491232}
\BIBentrySTDinterwordspacing

\bibitem{bhardwaj2014datahub}
A.~Bhardwaj, S.~Bhattacherjee, A.~Chavan, A.~Deshpande, A.~J. Elmore, S.~Madden, and A.~G. Parameswaran, ``Datahub: Collaborative data science \& dataset version management at scale,'' \emph{arXiv preprint arXiv:1409.0798}, 2014.

\bibitem{gonzalez2010google}
H.~Gonzalez, A.~Halevy, C.~S. Jensen, A.~Langen, J.~Madhavan, R.~Shapley, and W.~Shen, ``Google fusion tables: data management, integration and collaboration in the cloud,'' in \emph{Proceedings of the 1st ACM symposium on Cloud computing}, 2010, pp. 175--180.

\bibitem{winn2013open}
J.~Winn, ``Open data and the academy: An evaluation of ckan for research data management,'' 2013.

\bibitem{QuandlAl73:online}
``Quandl - alternativedata,'' \url{https://alternativedata.org/data_provider/quandl/}, (Accessed on 06/18/2024).

\bibitem{DataMark55:online}
``Data market :: There is a way to manage technology,'' \url{https://www.datamarket.com.tr/en/}, (Accessed on 06/18/2024).

\bibitem{KaggleYo44:online}
``Kaggle: Your machine learning and data science community,'' \url{https://www.kaggle.com/}, (Accessed on 06/18/2024).

\bibitem{john2017data}
T.~John and P.~Misra, \emph{Data lake for enterprises}.\hskip 1em plus 0.5em minus 0.4em\relax Packt Publishing Ltd, 2017.

\bibitem{vcuvs2022data}
B.~{\v{C}}u{\v{s}} and D.~Golec, ``Data lakehouse: Benefits in small and medium enterprises,'' \emph{Mednarodno inovativno poslovanje= Journal of Innovative Business and Management}, vol.~14, no.~2, pp. 1--10, 2022.

\bibitem{gupta2018practical}
S.~Gupta and V.~Giri, \emph{Practical Enterprise Data Lake Insights: Handle Data-Driven Challenges in an Enterprise Big Data Lake}.\hskip 1em plus 0.5em minus 0.4em\relax Apress, 2018.

\bibitem{panwar2022blockchain}
A.~Panwar, V.~Bhatnagar, M.~Khari, A.~W. Salehi, and G.~Gupta, ``A blockchain framework to secure personal health record (phr) in ibm cloud-based data lake,'' \emph{Computational Intelligence and Neuroscience}, vol. 2022, no.~1, p. 3045107, 2022.

\bibitem{halevy2016goods}
A.~Halevy, F.~Korn, N.~F. Noy, C.~Olston, N.~Polyzotis, S.~Roy, and S.~E. Whang, ``Goods: Organizing google's datasets,'' in \emph{Proceedings of the 2016 International Conference on Management of Data}, 2016, pp. 795--806.

\bibitem{sawadogo2021data}
P.~Sawadogo and J.~Darmont, ``On data lake architectures and metadata management,'' \emph{Journal of Intelligent Information Systems}, vol.~56, no.~1, pp. 97--120, 2021.

\bibitem{mami2019uniform}
M.~N. Mami, D.~Graux, S.~Scerri, H.~Jabeen, S.~Auer, and J.~Lehmann, ``Uniform access to multiform data lakes using semantic technologies,'' in \emph{Proceedings of the 21st International Conference on Information Integration and Web-based Applications \& Services}, 2019, pp. 313--322.

\bibitem{cafarella2018ten}
M.~Cafarella, A.~Halevy, H.~Lee, J.~Madhavan, C.~Yu, D.~Z. Wang, and E.~Wu, ``Ten years of webtables,'' \emph{Proceedings of the VLDB Endowment}, vol.~11, no.~12, pp. 2140--2149, 2018.

\bibitem{stonebraker2018c}
M.~Stonebraker, D.~J. Abadi, A.~Batkin, X.~Chen, M.~Cherniack, M.~Ferreira, E.~Lau, A.~Lin, S.~Madden, E.~O'Neil \emph{et~al.}, ``C-store: a column-oriented dbms,'' in \emph{Making Databases Work: the Pragmatic Wisdom of Michael Stonebraker}, 2018, pp. 491--518.

\bibitem{ritze2016profiling}
D.~Ritze, O.~Lehmberg, Y.~Oulabi, and C.~Bizer, ``Profiling the potential of web tables for augmenting cross-domain knowledge bases,'' in \emph{Proceedings of the 25th international conference on world wide web}, 2016, pp. 251--261.

\bibitem{church2017word2vec}
K.~W. Church, ``Word2vec,'' \emph{Natural Language Engineering}, vol.~23, no.~1, pp. 155--162, 2017.

\bibitem{pennington2014glove}
J.~Pennington, R.~Socher, and C.~D. Manning, ``Glove: Global vectors for word representation,'' in \emph{Proceedings of the 2014 conference on empirical methods in natural language processing (EMNLP)}, 2014, pp. 1532--1543.

\bibitem{trucscua2019efficiency}
M.~M. Tru{\c{s}}c{\u{a}}, ``Efficiency of svm classifier with word2vec and doc2vec models,'' in \emph{Proceedings of the International Conference on Applied Statistics}, vol.~1, no.~1, 2019, pp. 496--503.

\bibitem{wang2017machine}
Z.~Wang, ``Machine learning methods for finding textual features of depression from publications,'' 2017.

\bibitem{cao2020open}
E.~Cao, D.~Wang, J.~Huang, and W.~Hu, ``Open knowledge enrichment for long-tail entities,'' in \emph{Proceedings of The Web Conference 2020}, 2020, pp. 384--394.

\bibitem{sun2019gathering}
W.~Sun and N.~Wang, ``Gathering information on the web by consistent entity augmentation,'' \emph{Computing and Informatics}, vol.~38, no.~5, pp. 1039--1066, 2019.

\bibitem{yakout2012infogather}
M.~Yakout, K.~Ganjam, K.~Chakrabarti, and S.~Chaudhuri, ``Infogather: entity augmentation and attribute discovery by holistic matching with web tables,'' in \emph{Proceedings of the 2012 ACM SIGMOD International Conference on Management of Data}, 2012, pp. 97--108.

\bibitem{hamid2009data}
J.~S. Hamid, P.~Hu, N.~M. Roslin, V.~Ling, C.~M. Greenwood, and J.~Beyene, ``Data integration in genetics and genomics: methods and challenges,'' \emph{Human genomics and proteomics: HGP}, vol. 2009, 2009.

\bibitem{li2021deep}
Y.~Li, J.~Li, Y.~Suhara, J.~Wang, W.~Hirota, and W.-C. Tan, ``Deep entity matching: Challenges and opportunities,'' \emph{Journal of Data and Information Quality (JDIQ)}, vol.~13, no.~1, pp. 1--17, 2021.

\bibitem{10.1145/3431816}
\BIBentryALTinterwordspacing
------, ``Deep entity matching: Challenges and opportunities,'' \emph{J. Data and Information Quality}, vol.~13, no.~1, jan 2021. [Online]. Available: \url{https://doi.org/10.1145/3431816}
\BIBentrySTDinterwordspacing

\bibitem{10.1145/3318464.3389743}
\BIBentryALTinterwordspacing
R.~Wu, S.~Chaba, S.~Sawlani, X.~Chu, and S.~Thirumuruganathan, ``Zeroer: Entity resolution using zero labeled examples,'' in \emph{Proceedings of the 2020 ACM SIGMOD International Conference on Management of Data}, ser. SIGMOD '20.\hskip 1em plus 0.5em minus 0.4em\relax New York, NY, USA: Association for Computing Machinery, 2020, p. 1149–1164. [Online]. Available: \url{https://doi.org/10.1145/3318464.3389743}
\BIBentrySTDinterwordspacing

\bibitem{JAYARATNE2019996}
\BIBentryALTinterwordspacing
M.~Jayaratne, D.~Nallaperuma, D.~{De Silva}, D.~Alahakoon, B.~Devitt, K.~E. Webster, and N.~Chilamkurti, ``A data integration platform for patient-centered e-healthcare and clinical decision support,'' \emph{Future Generation Computer Systems}, vol.~92, pp. 996--1008, 2019. [Online]. Available: \url{https://www.sciencedirect.com/science/article/pii/S0167739X17308142}
\BIBentrySTDinterwordspacing

\bibitem{10.1007/978-3-319-07443-6_40}
B.~K{\"a}mpgen, T.~Weller, S.~O'Riain, C.~Weber, and A.~Harth, ``Accepting the xbrl challenge with linked data for financial data integration,'' in \emph{The Semantic Web: Trends and Challenges}, V.~Presutti, C.~d'Amato, F.~Gandon, M.~d'Aquin, S.~Staab, and A.~Tordai, Eds.\hskip 1em plus 0.5em minus 0.4em\relax Cham: Springer International Publishing, 2014, pp. 595--610.

\bibitem{10.14778/3229863.3240491}
\BIBentryALTinterwordspacing
R.~J. Miller, ``Open data integration,'' \emph{Proc. VLDB Endow.}, vol.~11, no.~12, p. 2130–2139, aug 2018. [Online]. Available: \url{https://doi.org/10.14778/3229863.3240491}
\BIBentrySTDinterwordspacing

\bibitem{10.1145/2882903.2912574}
\BIBentryALTinterwordspacing
X.~Chu, I.~F. Ilyas, S.~Krishnan, and J.~Wang, ``Data cleaning: Overview and emerging challenges,'' in \emph{Proceedings of the 2016 International Conference on Management of Data}, ser. SIGMOD '16.\hskip 1em plus 0.5em minus 0.4em\relax New York, NY, USA: Association for Computing Machinery, 2016, p. 2201–2206. [Online]. Available: \url{https://doi.org/10.1145/2882903.2912574}
\BIBentrySTDinterwordspacing

\bibitem{karlaš2020nearest}
B.~Karlaš, P.~Li, R.~Wu, N.~M. Gürel, X.~Chu, W.~Wu, and C.~Zhang, ``Nearest neighbor classifiers over incomplete information: From certain answers to certain predictions,'' 2020.

\bibitem{7474370}
H.~Liu, A.~Kumar~T.K., J.~P. Thomas, and X.~Hou, ``Cleaning framework for bigdata: An interactive approach for data cleaning,'' in \emph{2016 IEEE Second International Conference on Big Data Computing Service and Applications (BigDataService)}, 2016, pp. 174--181.

\bibitem{6816736}
Y.~Tong, C.~C. Cao, C.~J. Zhang, Y.~Li, and L.~Chen, ``Crowdcleaner: Data cleaning for multi-version data on the web via crowdsourcing,'' in \emph{2014 IEEE 30th International Conference on Data Engineering}, 2014, pp. 1182--1185.

\bibitem{9835412}
S.~Hao, P.~Li, R.~Wu, and X.~Chu, ``A model-agnostic approach for learning with noisy labels of arbitrary distributions,'' in \emph{2022 IEEE 38th International Conference on Data Engineering (ICDE)}, 2022, pp. 1219--1231.

\bibitem{1410106}
J.~Madhavan, P.~Bernstein, A.~Doan, and A.~Halevy, ``Corpus-based schema matching,'' in \emph{21st International Conference on Data Engineering (ICDE'05)}, 2005, pp. 57--68.

\bibitem{AmazonMe93:online}
``Amazon mechanical turk,'' \url{https://www.mturk.com/}, (Accessed on 06/18/2024).

\bibitem{10.1145/1866029.1866040}
\BIBentryALTinterwordspacing
G.~Little, L.~B. Chilton, M.~Goldman, and R.~C. Miller, ``Turkit: human computation algorithms on mechanical turk,'' in \emph{Proceedings of the 23nd Annual ACM Symposium on User Interface Software and Technology}, ser. UIST '10.\hskip 1em plus 0.5em minus 0.4em\relax New York, NY, USA: Association for Computing Machinery, 2010, p. 57–66. [Online]. Available: \url{https://doi.org/10.1145/1866029.1866040}
\BIBentrySTDinterwordspacing

\bibitem{10.1145/2398857.2384663}
\BIBentryALTinterwordspacing
D.~W. Barowy, C.~Curtsinger, E.~D. Berger, and A.~McGregor, ``Automan: a platform for integrating human-based and digital computation,'' \emph{SIGPLAN Not.}, vol.~47, no.~10, p. 639–654, oct 2012. [Online]. Available: \url{https://doi.org/10.1145/2398857.2384663}
\BIBentrySTDinterwordspacing

\bibitem{10.1145/2396761.2398421}
\BIBentryALTinterwordspacing
A.~G. Parameswaran, H.~Park, H.~Garcia-Molina, N.~Polyzotis, and J.~Widom, ``Deco: declarative crowdsourcing,'' in \emph{Proceedings of the 21st ACM International Conference on Information and Knowledge Management}, ser. CIKM '12.\hskip 1em plus 0.5em minus 0.4em\relax New York, NY, USA: Association for Computing Machinery, 2012, p. 1203–1212. [Online]. Available: \url{https://doi.org/10.1145/2396761.2398421}
\BIBentrySTDinterwordspacing

\bibitem{10.1145/3422622}
\BIBentryALTinterwordspacing
I.~Goodfellow, J.~Pouget-Abadie, M.~Mirza, B.~Xu, D.~Warde-Farley, S.~Ozair, A.~Courville, and Y.~Bengio, ``Generative adversarial networks,'' \emph{Commun. ACM}, vol.~63, no.~11, p. 139–144, oct 2020. [Online]. Available: \url{https://doi.org/10.1145/3422622}
\BIBentrySTDinterwordspacing

\bibitem{ARMANIOUS2020101684}
\BIBentryALTinterwordspacing
K.~Armanious, C.~Jiang, M.~Fischer, T.~Küstner, T.~Hepp, K.~Nikolaou, S.~Gatidis, and B.~Yang, ``Medgan: Medical image translation using gans,'' \emph{Computerized Medical Imaging and Graphics}, vol.~79, p. 101684, 2020. [Online]. Available: \url{https://www.sciencedirect.com/science/article/pii/S0895611119300990}
\BIBentrySTDinterwordspacing

\bibitem{pmlr-v157-zhao21a}
\BIBentryALTinterwordspacing
Z.~Zhao, A.~Kunar, R.~Birke, and L.~Y. Chen, ``Ctab-gan: Effective table data synthesizing,'' in \emph{Proceedings of The 13th Asian Conference on Machine Learning}, ser. Proceedings of Machine Learning Research, V.~N. Balasubramanian and I.~Tsang, Eds., vol. 157.\hskip 1em plus 0.5em minus 0.4em\relax PMLR, 17--19 Nov 2021, pp. 97--112. [Online]. Available: \url{https://proceedings.mlr.press/v157/zhao21a.html}
\BIBentrySTDinterwordspacing

\bibitem{JANSSEN2013123}
\BIBentryALTinterwordspacing
H.~Janssen, ``Monte-carlo based uncertainty analysis: Sampling efficiency and sampling convergence,'' \emph{Reliability Engineering System Safety}, vol. 109, pp. 123--132, 2013. [Online]. Available: \url{https://www.sciencedirect.com/science/article/pii/S0951832012001536}
\BIBentrySTDinterwordspacing

\bibitem{10.1145/130385.130417}
\BIBentryALTinterwordspacing
H.~S. Seung, M.~Opper, and H.~Sompolinsky, ``Query by committee,'' in \emph{Proceedings of the Fifth Annual Workshop on Computational Learning Theory}, ser. COLT '92.\hskip 1em plus 0.5em minus 0.4em\relax New York, NY, USA: Association for Computing Machinery, 1992, p. 287–294. [Online]. Available: \url{https://doi.org/10.1145/130385.130417}
\BIBentrySTDinterwordspacing

\bibitem{10.1145/3534678.3539476}
\BIBentryALTinterwordspacing
Y.~Kim and B.~Shin, ``In defense of core-set: A density-aware core-set selection for active learning,'' in \emph{Proceedings of the 28th ACM SIGKDD Conference on Knowledge Discovery and Data Mining}, ser. KDD '22.\hskip 1em plus 0.5em minus 0.4em\relax New York, NY, USA: Association for Computing Machinery, 2022, p. 804–812. [Online]. Available: \url{https://doi.org/10.1145/3534678.3539476}
\BIBentrySTDinterwordspacing

\bibitem{kapoor2007selective}
\BIBentryALTinterwordspacing
A.~Kapoor, E.~Horvitz, and S.~Basu, ``Selective supervision: Guiding supervised learning with decision-theoretic active learning,'' in \emph{IJCAI'07 Proceedings of the 20th international joint conference on Artifical intelligence}.\hskip 1em plus 0.5em minus 0.4em\relax Morgan Kaufmann Publishers Inc., January 2007, pp. 877--882. [Online]. Available: \url{https://www.microsoft.com/en-us/research/publication/selective-supervision-guiding-supervised-learning-decision-theoretic-active-learning/}
\BIBentrySTDinterwordspacing

\bibitem{10.1145/3025453.3026044}
\BIBentryALTinterwordspacing
J.~C. Chang, S.~Amershi, and E.~Kamar, ``Revolt: Collaborative crowdsourcing for labeling machine learning datasets,'' in \emph{Proceedings of the 2017 CHI Conference on Human Factors in Computing Systems}, ser. CHI '17.\hskip 1em plus 0.5em minus 0.4em\relax New York, NY, USA: Association for Computing Machinery, 2017, p. 2334–2346. [Online]. Available: \url{https://doi.org/10.1145/3025453.3026044}
\BIBentrySTDinterwordspacing

\bibitem{wang2012crowder}
J.~Wang, T.~Kraska, M.~J. Franklin, and J.~Feng, ``Crowder: Crowdsourcing entity resolution,'' 2012.

\bibitem{10.1145/1989323.1989486}
\BIBentryALTinterwordspacing
A.~Marcus, E.~Wu, D.~R. Karger, S.~Madden, and R.~C. Miller, ``Demonstration of qurk: a query processor for humanoperators,'' in \emph{Proceedings of the 2011 ACM SIGMOD International Conference on Management of Data}, ser. SIGMOD '11.\hskip 1em plus 0.5em minus 0.4em\relax New York, NY, USA: Association for Computing Machinery, 2011, p. 1315–1318. [Online]. Available: \url{https://doi.org/10.1145/1989323.1989486}
\BIBentrySTDinterwordspacing

\bibitem{10.1145/3035918.3056442}
\BIBentryALTinterwordspacing
A.~J. Ratner, S.~H. Bach, H.~R. Ehrenberg, and C.~R\'{e}, ``Snorkel: Fast training set generation for information extraction,'' in \emph{Proceedings of the 2017 ACM International Conference on Management of Data}, ser. SIGMOD '17.\hskip 1em plus 0.5em minus 0.4em\relax New York, NY, USA: Association for Computing Machinery, 2017, p. 1683–1686. [Online]. Available: \url{https://doi.org/10.1145/3035918.3056442}
\BIBentrySTDinterwordspacing

\bibitem{Wu_2021}
\BIBentryALTinterwordspacing
R.~Wu, P.~Sakala, P.~Li, X.~Chu, and Y.~He, ``Demonstration of panda: a weakly supervised entity matching system,'' \emph{Proceedings of the VLDB Endowment}, vol.~14, no.~12, p. 2735–2738, Jul. 2021. [Online]. Available: \url{http://dx.doi.org/10.14778/3476311.3476332}
\BIBentrySTDinterwordspacing

\bibitem{zhang2022survey}
J.~Zhang, C.-Y. Hsieh, Y.~Yu, C.~Zhang, and A.~Ratner, ``A survey on programmatic weak supervision,'' 2022.

\bibitem{10.1145/3191513}
\BIBentryALTinterwordspacing
T.~Mitchell, W.~Cohen, E.~Hruschka, P.~Talukdar, B.~Yang, J.~Betteridge, A.~Carlson, B.~Dalvi, M.~Gardner, B.~Kisiel, J.~Krishnamurthy, N.~Lao, K.~Mazaitis, T.~Mohamed, N.~Nakashole, E.~Platanios, A.~Ritter, M.~Samadi, B.~Settles, R.~Wang, D.~Wijaya, A.~Gupta, X.~Chen, A.~Saparov, M.~Greaves, and J.~Welling, ``Never-ending learning,'' \emph{Commun. ACM}, vol.~61, no.~5, p. 103–115, apr 2018. [Online]. Available: \url{https://doi.org/10.1145/3191513}
\BIBentrySTDinterwordspacing

\bibitem{10.1145/2623330.2623623}
\BIBentryALTinterwordspacing
X.~Dong, E.~Gabrilovich, G.~Heitz, W.~Horn, N.~Lao, K.~Murphy, T.~Strohmann, S.~Sun, and W.~Zhang, ``Knowledge vault: a web-scale approach to probabilistic knowledge fusion,'' in \emph{Proceedings of the 20th ACM SIGKDD International Conference on Knowledge Discovery and Data Mining}, ser. KDD '14.\hskip 1em plus 0.5em minus 0.4em\relax New York, NY, USA: Association for Computing Machinery, 2014, p. 601–610. [Online]. Available: \url{https://doi.org/10.1145/2623330.2623623}
\BIBentrySTDinterwordspacing

\bibitem{9417095}
H.~Chen, J.~Chen, and J.~Ding, ``Data evaluation and enhancement for quality improvement of machine learning,'' \emph{IEEE Transactions on Reliability}, vol.~70, no.~2, pp. 831--847, 2021.

\bibitem{NEURIPS2019_95323660}
\BIBentryALTinterwordspacing
A.~Kirsch, J.~van Amersfoort, and Y.~Gal, ``Batchbald: Efficient and diverse batch acquisition for deep bayesian active learning,'' in \emph{Advances in Neural Information Processing Systems}, H.~Wallach, H.~Larochelle, A.~Beygelzimer, F.~d\textquotesingle Alch\'{e}-Buc, E.~Fox, and R.~Garnett, Eds., vol.~32.\hskip 1em plus 0.5em minus 0.4em\relax Curran Associates, Inc., 2019. [Online]. Available: \url{https://proceedings.neurips.cc/paper_files/paper/2019/file/95323660ed2124450caaac2c46b5ed90-Paper.pdf}
\BIBentrySTDinterwordspacing

\bibitem{10.1145/3289600.3291035}
\BIBentryALTinterwordspacing
L.~Han, K.~Roitero, U.~Gadiraju, C.~Sarasua, A.~Checco, E.~Maddalena, and G.~Demartini, ``All those wasted hours: On task abandonment in crowdsourcing,'' in \emph{Proceedings of the Twelfth ACM International Conference on Web Search and Data Mining}, ser. WSDM '19.\hskip 1em plus 0.5em minus 0.4em\relax New York, NY, USA: Association for Computing Machinery, 2019, p. 321–329. [Online]. Available: \url{https://doi.org/10.1145/3289600.3291035}
\BIBentrySTDinterwordspacing

\end{thebibliography}

\end{document}